\documentclass[12pt]{article}
\usepackage{bbm,latexsym}
\usepackage{epsfig}
\newcommand{\be}{\begin{equation}}
\newcommand{\ee}{\end{equation}}
\newcommand{\ba}{\begin{eqnarray}}
\newcommand{\ea}{\end{eqnarray}}

\renewcommand{\thefootnote}{\fnsymbol{footnote}}

\newcommand{\gtrsim}{\:\mbox{\raisebox{-4pt}{$\stackrel%
{\displaystyle >}{\sim}$}}\:}

\newcommand{\mnu}{\mathcal{M}_\nu}
\newcommand{\s}{\hspace{0.5mm}}
\begin{document}
\title{\normalsize \hfill UWThPh-2011-14 \\[1cm] \LARGE
Maximal atmospheric neutrino mixing from texture zeros and
quasi-degenerate neutrino masses}
\author{W. Grimus\thanks{E-mail: walter.grimus@univie.ac.at} 
\setcounter{footnote}{6}
and P.O. Ludl\thanks{E-mail: patrick.ludl@univie.ac.at} \\[4mm]
\small University of Vienna, Faculty of Physics \\
\small Boltzmanngasse 5, A--1090 Vienna, Austria \\[4.6mm]}

\date{21 April 2011}

\maketitle

\begin{abstract}
It is well-known that, 
in the basis where the charged-lepton mass matrix is diagonal,
there are seven cases 
of two texture zeros in Majorana neutrino mass matrices
that are compatible with all experimental data. 
We show that two of these cases, 
namely B$_3$ and B$_4$ in the classification of 
Frampton, Glashow and Marfatia, are special in the sense 
that they automatically lead to near-maximal atmospheric neutrino mixing in
the limit of a quasi-degenerate neutrino mass spectrum. 
This property holds true irrespective of the values of the solar
and reactor mixing angles because, 
for these two cases, in the limit of a quasi-degenerate spectrum, 
the second and third row of the lepton mixing matrix are, up to
signs, approximately complex-conjugate to each other. 
Moreover, in the same
limit the aforementioned cases also develop a maximal CP-violating CKM-type
phase, provided the reactor mixing angle is not too small. 
\end{abstract}

\newpage


\setcounter{footnote}{0}
\renewcommand{\thefootnote}{\arabic{footnote}}

\section{Introduction}

It is by now well-established that at least two of the neutrino masses
$m_j$ ($j=1,2,3$) are non-zero. The same applies to the angles in the lepton
mixing matrix $V$. 
Its parameterization is usually chosen in analogy to the
CKM matrix as~\cite{rpp}
\be\label{V}
V = \left( \begin{array}{ccc}
c_{13} c_{12} &
c_{13} s_{12} &
s_{13} e^{-i \delta} \\
- c_{23} s_{12} - s_{23} s_{13} c_{12} e^{i \delta} &
c_{23} c_{12} - s_{23} s_{13} s_{12} e^{i \delta} &
s_{23} c_{13} \\
s_{23} s_{12} - c_{23} s_{13} c_{12} e^{i \delta} &
-s_{23} c_{12} - c_{23} s_{13} s_{12} e^{i \delta} &
c_{23} c_{13} 
\end{array} \right),
\ee
with $c_{ij} \equiv \cos{\theta_{ij}}$
and $s_{ij} \equiv \sin{\theta_{ij}}$, 
the $\theta_{ij}$ being angles of the first quadrant.
While the angles $\theta_{12}$ and $\theta_{23}$ are approximately 
$34^\circ$ and $45^\circ$, respectively,
the angle $\theta_{13}$ is compatible with
zero~\cite{results,schwetz}. All data on lepton mixing are compatible
with the tri-bimaximal matrix
\begin{equation}
V_\mathrm{HPS} \equiv \left( \begin{array}{rrc}
2/\sqrt{6} & 1/\sqrt{3} & 0 \\ 
-1/\sqrt{6} & 1/\sqrt{3} & 1/\sqrt{2} \\ 
1/\sqrt{6} & -1/\sqrt{3} & 1/\sqrt{2}
\end{array} \right),
\label{HPS}
\end{equation}
which has been put forward by Harrison,
Perkins and Scott (HPS)~\cite{HPS} already in 2002.

Equation~(\ref{HPS}) has lead to the speculation that there is a
\emph{non-abelian} family symmetry behind the
scenes,\footnote{However, recently, it has been argued that
  tri-bimaximal mixing might nevertheless have an accidental
  origin~\cite{abbas}.} enforcing 
$s_{23}^2 = 1/2$ in particular. This speculation
is in accord with the finding of~\cite{low} that the only extremal angle
which can be obtained by an \emph{abelian} symmetry is $\theta_{13} =
0^\circ$, \textit{i.e.}, $\theta_{23} = 45^\circ$ cannot be enforced
by an abelian symmetry. A favourite non-abelian group in this context
is $A_4$~\cite{A4}. For recent developments and other favourite groups 
see the reviews in~\cite{review} and references therein, for attempts
on systematic studies see~\cite{tanimoto,wingerter,plentinger} (the latter
paper refers to abelian symmetries).

However, there is an alternative to non-abelian groups. It is not necessary
that, for instance, $\theta_{23} = 45^\circ$ is \emph{exactly} realized at
some energy scale. It suffices that such a relation is fulfilled with
reasonable accuracy. This could happen without need for a non-abelian
symmetry. In order to pin down what we mean specifically we consider the
structure of the mixing matrix $V$. It has two contributions, the unitary
matrices $U_\ell$ and $U_\nu$, stemming from the diagonalization of the
charged-lepton mass matrix $M_\ell$ and of the neutrino mass matrix $\mnu$,
respectively. Then the matrix 
\be
U \equiv U_\ell^\dagger U_\nu = e^{i\hat \alpha}\, V \,
e^{i\hat \sigma}
\ee
occurs in the charged-current interaction and the lepton mixing matrix
$V$ defined above is obtained by removing the diagonal unitary
matrices 
\be
e^{i\hat \alpha} = \mbox{diag} \left( 
e^{i\alpha_1},\,  e^{i\alpha_2},\,  e^{i\alpha_3} \right) 
\quad \mbox{and} \quad 
e^{i\hat \sigma}  = \mbox{diag} \left( 
e^{i\sigma_1},\,  e^{i\sigma_2},\,  e^{i\sigma_3} \right) 
\ee
from $U$. Without loss of generality we will use the convention 
$e^{i\sigma_3} = 1$ in the following.
Suppose that we have a model in which $U_\ell$ and $U_\nu$ are functions of
the charged-lepton and neutrino mass ratios, respectively, and that
these mass ratios also parameterize the deviations of $U_\ell$ and
$U_\nu$ from a diagonal form. In $U_\ell$ these
ratios are $m_e/m_\mu$, $m_e/m_\tau$ and $m_\mu/m_\tau$.
Since the mass hierarchy in the charged-lepton sector is rather
strong, $U_\ell$ is approximately a diagonal matrix of phase
factors, with the possible exception of the occurrence of $m_\mu/m_\tau$; if
this ratio appears in a square root 
in analogy to the famous formula 
$\sin\theta_c \simeq \sqrt{m_d/m_s}$ for the Cabibbo angle~\cite{gatto},
with quark masses $m_d$ and $m_s$,
then $\sqrt{m_\mu/m_\tau} \simeq 0.24$ is even larger than 
$\sin\theta_c$. The simplest way to avoid such a deviation of $U_\ell$ from 
a diagonal matrix is to have a model which, through its symmetries, enforces a
diagonal $M_\ell$. Switching to $U_\nu$, we point out that up to now the type of
neutrino mass spectrum is completely unknown~\cite{rpp}. A particularly exciting
possibility would be a quasi-degenerate mass spectrum in which case the
neutrino mass ratios could be close to one such that effectively $U_\nu$ is
independent of the masses and could look like a matrix of pure numbers,
potentially disturbed by phase factors. 
Thus, with $U_\ell$ sufficiently close to a diagonal matrix and a
quasi-degenerate neutrino mass spectrum it might be possible to simulate a
mixing matrix $V$ consisting of ``pure numbers,'' leading for instance to an
atmospheric neutrino mixing angle $\theta_{23}$ which is in practice
indistinguishable from $45^\circ$.

The advantage is that such a scenario could be achieved with texture zeros 
and that texture zeros in mass matrices may \emph{always} be explained
by abelian symmetries~\cite{GJLT},
at the expense of an extended scalar sector in renormalizable 
models.\footnote{We emphasize that our approach is 
  different from that of~\cite{plentinger} where the Froggatt--Nielsen
  mechanism~\cite{froggatt} is used and, therefore, order-of-magnitude
  relations among the elements of mass matrices are assumed.}

Let us summarize the assumptions of this paper:
\begin{itemize}
\item
$U_\ell$ is sufficiently close to a diagonal matrix such that in good
  approximation it does not contribute to $V$. 
\item 
The neutrino mass spectrum is quasi-degenerate.
\item
The symmetry groups we have in mind are abelian, \textit{i.e.}, we deal with 
texture zeros.
\end{itemize}
In the following we will show that these assumptions can indeed lead to
a realization of maximal atmospheric neutrino mixing, in the 
framework which consists of Majorana neutrino mass matrices with two texture
zeros and a diagonal mass matrix $M_\ell$; two of the viable cases of neutrino
mass matrices classified in~\cite{FGM} exhibit precisely the desired features. 

In section~\ref{framework} we review the viable textures presented
in~\cite{FGM} and point out models in which they
can be realized. Then, in section~\ref{B}, we discuss the
phenomenology of the cases B$_3$ and B$_4$ of~\cite{FGM} 
in the light of the philosophy specified above.
The remaining cases are discussed in 
section~\ref{remaining}. We summarize our findings 
in section~\ref{concl}.

\section{The framework}
\label{framework}

Assuming the neutrinos to be Majorana particles, 
the neutrino mass term is given by
\begin{equation}
\mathcal{L}_{\nu\, \mathrm{mass}}
= \frac{1}{2}\, \nu_\mathrm{L}^\mathrm{T} C^{-1}
\mathcal{M}_\nu \nu_\mathrm{L}
+ \mbox{H.c.},
\end{equation}
with a symmetric mass matrix $\mathcal{M}_\nu$.
In the basis where the charged-lepton mass matrix is diagonal,
there are seven possibilities
for an $\mathcal{M}_\nu$ with two texture zeros
which are compatible with all available neutrino data,
as was shown in~\cite{FGM}.
These seven viable cases are listed in table~\ref{tab}.
The phenomenology of those seven mass matrices has been
discussed in~\cite{FGM,xing1,xing2}. Moreover, case~C 
has also been investigated in~\cite{GL05}.
\begin{table}
\renewcommand{\arraystretch}{1.2}
\begin{center}
\begin{tabular}{c|c}
case & texture zeros \\ \hline
A$_1$ & $(\mnu)_{ee} = (\mnu)_{e\mu} = 0$ \\
A$_2$ & $(\mnu)_{ee} = (\mnu)_{e\tau} = 0$ \\
B$_1$ & $(\mnu)_{\mu\mu} = (\mnu)_{e\tau} = 0$ \\
B$_2$ & $(\mnu)_{\tau\tau} = (\mnu)_{e\mu} = 0$ \\
B$_3$ & $(\mnu)_{\mu\mu} = (\mnu)_{e\mu} = 0$ \\
B$_4$ & $(\mnu)_{\tau\tau} = (\mnu)_{e\tau} = 0$ \\
C     & $(\mnu)_{\mu\mu} = (\mnu)_{\tau\tau} = 0$
\end{tabular}
\end{center}
\caption{The viable cases in the framework of two zeros in the
  Majorana neutrino mass matrix 
  $\mnu$ and a diagonal charged-lepton mass matrix
  $M_\ell$~\cite{FGM}. \label{tab}} 
\end{table}

There are several ways to construct models where the cases of table~\ref{tab}
together with a diagonal charged-lepton mass matrix are realized by
symmetries. Five of the seven mass matrices, but not B$_1$ and B$_2$, 
have various embeddings in the seesaw mechanism~\cite{seesaw},
by placing zeros in the Majorana mass matrix $M_\mathrm{R}$
of the right-handed neutrino singlets $\nu_\mathrm{R}$
and in the Dirac mass matrix $M_\mathrm{D}$
connecting the $\nu_\mathrm{R}$
with the $\nu_\mathrm{L}$~\cite{kageyama}. 
With the methods described in~\cite{GJLT}, one can then construct
models where the zeros in the various mass matrices, including the six
off-diagonal zeros in $M_\ell$, are enforced by abelian symmetries.

Four of the seven textures of table~\ref{tab}, 
namely A$_1$, A$_2$, B$_3$, B$_4$, have a realization in the seesaw
mechanism with a \emph{diagonal} $M_D$~\cite{kageyama,lavoura}: by
suitably placing two texture zeros in $M_R$ or, equivalently, in $\mnu^{-1}$,
these four textures can be obtained.\footnote{There are three more
  viable cases of texture zeros in $\mnu^{-1}$ which do not correspond
  to texture zeros in $\mnu$~\cite{lavoura}.}
Actually, now we are dealing with 14 texture zeros, namely six in
$M_\ell$ and $M_D$ each and two in $M_R$.
In order to construct models for these four cases, one Higgs doublet
is sufficient, but one needs two scalar gauge singlets in order to
implement the desired form of $M_R$~\cite{lavoura}.

\emph{All} of the seven cases of table~\ref{tab} can be realized as
models in scalar-triplet extensions of the Standard Model~\cite{GL05},
\textit{i.e.}, in the type~II seesaw mechanism~\cite{seesawII} without
any right-handed neutrino singlets.

\section{Cases B$_3$ and B$_4$}
\label{B}

In this section we discuss the cases B$_3$ and B$_4$ which
correspond to the Majorana mass matrices
\begin{equation}\label{B3}
\mbox{B}_3: \quad
\mathcal{M}_\nu \sim 
\left( \begin{array}{ccc} 
\times & 0 & \times \\ 0 & 0 & \times \\ \times & \times & \times 
\end{array} \right),
\qquad
\mbox{B}_4: \quad
\mathcal{M}_\nu \sim 
\left( \begin{array}{ccc} 
\times & \times & 0 \\ \times & \times & \times \\ 0 & \times & 0 
\end{array} \right).
\end{equation}
The symbol $\times$ denotes non-zero matrix elements. 
The equations which follow from these cases have the form
\be\label{20}
\sum_{j=1}^3 V_{\alpha j} V_{\alpha j} \mu_j = 
\sum_{j=1}^3 V_{\alpha j} V_{\beta j} \mu_j = 0
\quad \mbox{with} \quad
\mu_j \equiv m_j e^{2i\sigma_j}
\ee
and $\alpha \neq \beta$, where B$_3$ is given by 
$(\alpha, \beta) = (\mu, e)$ and B$_4$ by 
$(\alpha, \beta) = (\tau, e)$.

Equation~(\ref{20}) can be considered 
from a linear-algebra perspective. Defining line vectors 
\be
z_\alpha = \left( V_{\alpha j} \right), \quad
z_\beta  = \left( V_{\beta j}  \right)
\ee
of $V$, equation~(\ref{20}) tells us that, because of the unitarity of
$V$, the line vector 
\be
\left( V_{\alpha 1}^*\, \mu_1^*,\, V_{\alpha 2}^*\, \mu_2^*,\, 
V_{\alpha 3}^*\, \mu_3^* \right)
\ee
is orthogonal to both,
$z_\alpha$ and $z_\beta$. Therefore, this vector must be proportional
to the line vector $z_\gamma$ with $\gamma \neq \alpha, \beta$, 
and we obtain 
\be\label{gamma}
z_\gamma = \left( V_{\gamma j}  \right) = 
\frac{e^{i\phi}}{N_\alpha} \left( V_{\alpha j}^*\, \mu_j^*
\right) \quad \mbox{with} \quad 
N_\alpha^2 = \sum_{k=1}^3 \left| V_{\alpha k} \right|^2 m_k^2.
\ee

We use equation~(\ref{gamma}) for the discussion
of the physical consequences of cases B$_3$ and B$_4$. 
We begin with case B$_3$ where $\gamma = \tau$.
Defining 
$\epsilon = s_{13} e^{i\delta}$, $t_{12} = s_{12}/c_{12}$ and 
$t_{23} = s_{23}/c_{23}$, from equations~(\ref{V}) and 
(\ref{gamma}) we find the following relations:
\be\label{basic3}
\mathrm{B}_3: \quad
\frac{\mu_1}{m_3} = 
\frac{V_{\mu 3} V_{\tau 1}^*}{V_{\mu 1} V_{\tau 3}^*} =
-\frac{t_{12} t_{23} - \epsilon^*}{t_{12} + t_{23} \epsilon}\, t_{23},
\quad
\frac{\mu_2}{m_3} = 
\frac{V_{\mu 3} V_{\tau 2}^*}{V_{\mu 2} V_{\tau 3}^*} =
-\frac{t_{23} + t_{12} \epsilon^*}{1 - t_{12} t_{23} \epsilon}\, t_{23}.
\ee
Alternatively, one can use the procedure of~\cite{xing1} to arrive at
these expressions. 

The analysis of equation~(\ref{basic3}) proceeds in the following
way. We define
\be
\rho_j = \left( \frac{m_j}{m_3} \right)^2 \quad (j=1,2),
\ee
take the absolute values of the two expressions in
equation~(\ref{basic3}) and eliminate 
\be\label{zeta}
\zeta \equiv 2 t_{12} t_{23} s_{13} \cos\delta =
\frac{t_{12}^2 t_{23}^2 + s_{13}^2 - \rho_1 (t_{12}^2 + t_{23}^2
  s_{13}^2)/t_{23}^2}{1 + \rho_1/t_{23}^2}.
\ee
Then we end up with a cubic equation for $t_{23}^2$:
\be\label{t23}
t_{23}^6 + t_{23}^4 \left[ s_{13}^2 + c_{13}^2 \left( c_{12}^2 \rho_1
  + s_{12}^2 \rho_2 \right) \right] -
t_{23}^2 \left[ s_{13}^2 \rho_1 \rho_2 + c_{13}^2 \left( s_{12}^2 \rho_1
  + c_{12}^2 \rho_2 \right) \right] - \rho_1 \rho_2 = 0.
\ee
Inspection of this equation shows that it has a unique positive
root. Thus, given the neutrino masses $m_1$, $m_2$, $m_3$ and the
mixing angles $\theta_{12}$ and $\theta_{13}$, equation~(\ref{t23})
determines $\theta_{23}$. Using equations~(\ref{basic3}) and
(\ref{zeta}), we adopt the following philosophy:
\be
\mbox{input:}\;\;  m_{1,2,3},\, \theta_{12},\, \theta_{13} 
\quad \Rightarrow \quad 
\mbox{predictions:} \;\; \theta_{23},\, \delta,\, 2\sigma_{1,2}.
\ee
Since the Majorana phases $2\sigma_{1,2}$ are not directly accessible
to experimental scrutiny, we will later consider instead the observable
$m_{\beta\beta}$, the effective mass in neutrinoless double-beta decay.

An approximate solution of equation~(\ref{t23}) for a quasi-degenerate
neutrino mass spectrum is given by 
\be\label{approxb3}
t_{23}^2 \simeq 1 - \frac{1}{2} \, 
\frac{\Delta m_{31}^2}{m_1^2} \left( 1 + s_{13}^2 \right)
\quad \mbox{or} \quad 
s_{23}^2 \simeq \frac{1}{2} - \frac{1}{8} \, 
\frac{\Delta m_{31}^2}{m_1^2} \left( 1 + s_{13}^2 \right),
\ee
where corrections of order  
$(\Delta m_{31}^2/m_1^2)^2$ and $\Delta m_{21}^2/m_1^2$ 
have been neglected.\footnote{Note that
  $\Delta m_{21}^2 = m_2^2 - m_1^2 > 0$ whereas 
  $\Delta m_{31}^2 = m_3^2 - m_1^2$ can have either sign: $\Delta
  m_{31}^2 > 0$ indicates the normal order of the neutrino mass
  spectrum and $\Delta m_{31}^2 < 0$ the inverted order.}
The latter term is small because we know from experiment that
$\Delta m_{21}^2/|\Delta m_{31}^2| \sim 1/30$~\cite{results,schwetz}.
We observe that the leading correction to $t_{23}^2$ is independent of
$s_{12}^2$. 

\begin{figure}[t]
\begin{center}
\epsfig{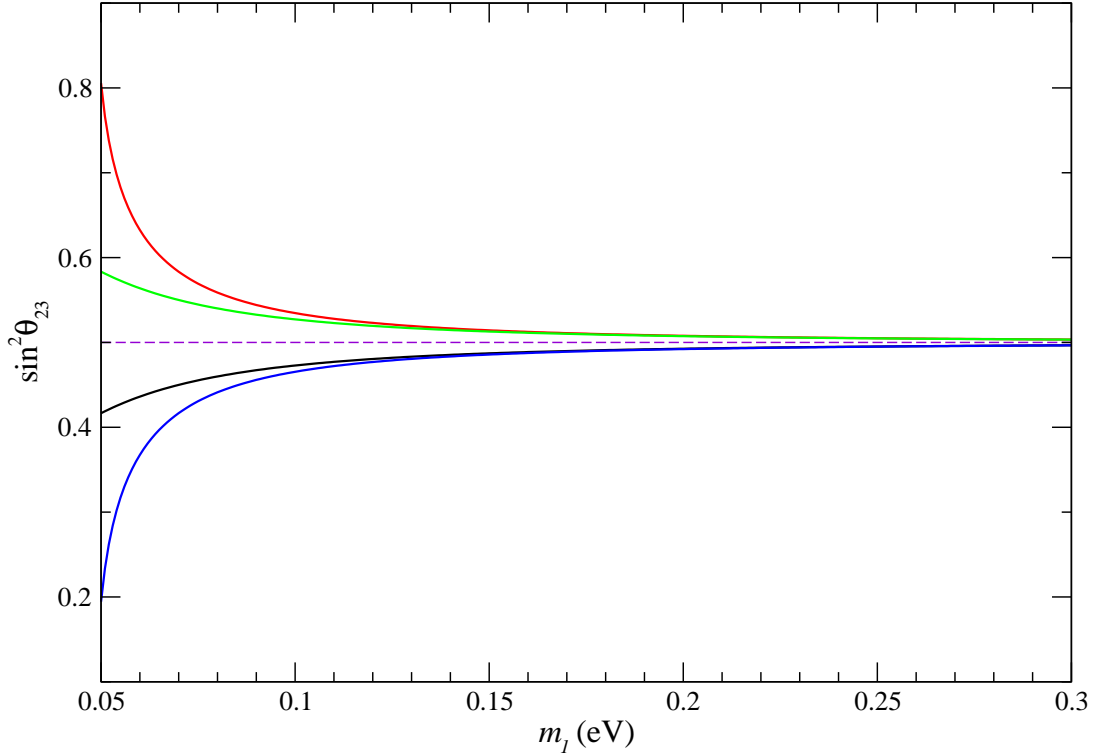}
\end{center}
\caption{$s_{23}^2$ as a function of $m_1$. In descending order the full curves
  refer to case B$_3$ (inverted spectrum), case B$_4$ (normal spectrum), 
  case B$_3$ (normal spectrum), and case B$_4$ (inverted spectrum). The dashed
  line indicates the value 0.5, \textit{i.e.}, maximal atmospheric mixing.
  In this plot, for $s_{12}^2$, $s_{13}^2$, $\Delta m^2_{21}$ and
  $\Delta m^2_{31}$ the best-fit values of~\cite{schwetz} have been used.
  \label{fig-theta23}}
\end{figure}

Case B$_4$ is treated analogously. We obtain 
\be\label{basic4}
\mathrm{B}_4: \quad
\frac{\mu_1}{m_3} = 
\frac{V_{\mu 1}^* V_{\tau 3}}{V_{\mu 3}^* V_{\tau 1}} =
-\frac{t_{12} + t_{23} \epsilon^*}{t_{12} t_{23} - \epsilon}\, 
\frac{1}{t_{23}},
\quad
\frac{\mu_2}{m_3} = 
\frac{V_{\mu 2}^* V_{\tau 3}}{V_{\mu 3}^* V_{\tau 2}} =
-\frac{1 - t_{12} t_{23} \epsilon^*}{t_{23} + t_{12} \epsilon}\, 
\frac{1}{t_{23}}.
\ee
Comparison with equation~(\ref{basic3}) shows that in the present case the
cubic equation for $t_{23}^2$ is obtained from equation~(\ref{t23}) by the
replacement $\rho_1 \to 1/\rho_1$, $\rho_2 \to 1/\rho_2$.
It is then easy to show that the atmospheric mixing angles in the cases
B$_3$ and B$_4$ are related by 
\be\label{3-4}
\left. t_{23}^2 \right|_{\mathrm{B}_4} = \left( \left. t_{23}^2
\right|_{\mathrm{B}_3} \right)^{-1}
\quad \mbox{or} \quad 
\left. s_{23}^2 \right|_{\mathrm{B}_4} = 1 - \left. s_{23}^2
\right|_{\mathrm{B}_3}.
\ee
Accordingly, the curves for B$_4$ in figure~\ref{fig-theta23} are
obtained from those of B$_3$ by reflection at the dashed line.

\begin{figure}[t]
\begin{center}
\epsfig{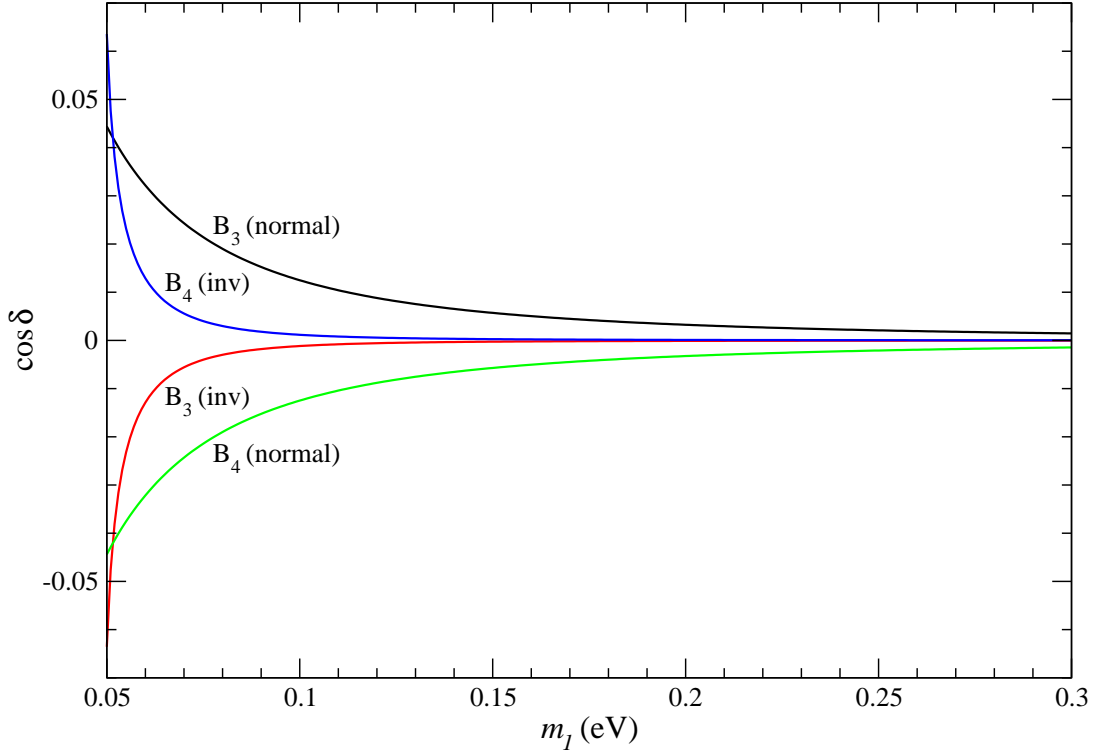}
\end{center}
\caption{$\cos \delta$ as a function of $m_1$. For further details see the
  legend of figure~\ref{fig-theta23}.
  \label{fig-cosd}}
\end{figure}
In figures~\ref{fig-theta23} and~\ref{fig-cosd} we have plotted
$s_{23}^2$ and $\cos \delta$ versus $m_1$, respectively, for cases
B$_3$ and B$_4$. 
For definiteness, for the solar and reactor mixing angles and the
mass-squared differences we have used the best-fit
values listed in~\cite{schwetz}:
$s_{12}^2 = 0.316$, $\Delta m^2_{21} = 7.64 \times 10^{-5}$\,eV$^2$,
which are the same values for both normal and inverted spectrum, and  
$s_{13}^2 = 0.017$, $\Delta m^2_{31} = 2.45 \times 10^{-3}$\,eV$^2$
for the normal and 
$s_{13}^2 = 0.020$, $\Delta m^2_{31} = -2.34 \times 10^{-3}$\,eV$^2$
for the inverted spectrum.
The two figures illustrate nicely that in all four instances (cases
B$_3$ and B$_4$ and both spectra) in the limit $m_1 \to \infty$ we
find $s_{23}^2 \to 1/2$ and $\cos \delta \to 0$.

Some remarks are at order. First of all, 
by a numerical comparison it turns out that
the approximate formula~(\ref{approxb3}) works
quite well. The deviation from the exact
value of $s_{23}^2$ is less than 3\% at $m_1 = 0.08$\,eV and the 
approximation rapidly improves at larger $m_1$. Secondly,
from equations~(\ref{basic3}) and~(\ref{basic4}) we read off that
cases B$_3$ and B$_4$ do not allow $s_{13} = 0$ because this would lead to
$\mu_1 = \mu_2$. However, this observation does not give a strong
restriction on $s_{13}$, as we find numerically. Thirdly, the lower
bound on $s_{13}$ is correlated with a lower bound on $m_1$. The reason
is that, in our treatment of cases B$_3$ and B$_4$, $\cos \delta$ is
computed via equation~(\ref{zeta}) after the determination of $s_{23}^2$ by
equation~(\ref{t23}); then the condition $|\cos \delta| \leq 1$
leads to the lower bound on $m_1$. For the inverted spectrum we obtain
numerically that the lower bound $m_1 \gtrsim 0.05$\,eV is rather
stable for $s_{13}^2 \gtrsim 0.0001$. The normal spectrum allows
smaller values of $m_1$, for instance, $m_1 \gtrsim 0.03$\,eV at 
$s_{13}^2 \simeq 0.0001$ and $m_1 \gtrsim 0.01$\,eV at 
$s_{13}^2 \simeq 0.01$. Therefore, cases B$_3$ and B$_4$ do \emph{not}
automatically entail a quasi-degenerate spectrum which would require
something like $m_1 > 0.1$\,eV. In accord with the philosophy of this
paper we really have to postulate such a spectrum and only for
quasi-degeneracy we obtain an atmospheric mixing angle sufficiently
close to $45^\circ$.

Computing an approximation for $\cos \delta$ is a bit laborious. It turns out
that, due to the smallness of $s_{13}^2$, 
it is necessary to expand
$\cos \delta$ to second order in both 
\be
\Delta_1=\frac{\Delta m_{31}^2}{m_1^2} \quad\mbox{and}\quad
\Delta_2=\frac{\Delta m_{21}^2}{m_1^2}, 
\ee
in order to obtain a reasonable accuracy. 
The result is\footnote{Note that the coefficient of
  $\Delta_1 \Delta_2$ is zero.}
\be\label{cosdeltaapprox}
\mathrm{cos}\s\delta\simeq\mp \frac{s_{13}t_{12}}{4} \left\{
\left(1-\frac{1}{t_{12}^2}\right)\left(\Delta_1-\frac{1}{2}\Delta_1^2\right)
+
\frac{s_{12}^2c_{13}^2-1}{s_{13}^2}\left(\Delta_2-\frac{1}{2}\Delta_2^2\right)
\right\}, 
\ee
where the minus and plus signs correspond to B$_3$ and B$_4$,
respectively. At $m_1=0.16$\,eV the approximation
(\ref{cosdeltaapprox}) deviates from the exact value by less than 1\%
(5\%) assuming a normal (inverted) spectrum. 
Actually, the sign difference in equation~(\ref{cosdeltaapprox})
between cases B$_3$ and B$_4$ holds to all orders; with
equations~(\ref{zeta}) and~(\ref{3-4}) it is easy to show that
\be
\left. \cos \delta \right|_{\mathrm{B}_4} = 
-\left. \cos \delta \right|_{\mathrm{B}_3},
\ee
in perfect agreement with the numerical computation.

The general formula for the effective mass in neutrinoless double-beta
decay (for reviews see for instance~\cite{mbb-review}) is given by 
the formula
\be
m_{\beta\beta} = \left| \left( \mnu \right)_{ee} \right| = 
m_3 \left| c_{13}^2 \left( c_{12}^2 \frac{\mu_1}{m_3} + s_{12}^2 
\frac{\mu_2}{m_3} \right) +
  \left( \epsilon^* \right)^2 \right|.
\ee
With $\mu_1$ and $\mu_2$ from equations~(\ref{basic3}) and
(\ref{basic4}) we specify it to cases B$_3$ and B$_4$, respectively.
Inspection of the same equations 
reveals a simple procedure to switch from B$_3$ to B$_4$:
\be
\left. \frac{\mu_j}{m_3} \right|_{\mathrm{B}_4} = 
\left( \left. \frac{m_3}{\mu_j} \right|_{\mathrm{B}_3} \right)^*
\quad (j=1,2).
\ee
In $\mu_1$ and $\mu_2$ we need to insert the numerical values obtained
for $t_{23}$ and $\delta$. 
Equation~(\ref{zeta}) determines $\cos \delta$, therefore, 
$\sin \delta$ is only determined up to a sign. However, since 
$\sin\delta \leftrightarrow -\sin\delta$ corresponds to 
$\epsilon \leftrightarrow \epsilon^*$, this sign has no effect 
on $m_{\beta\beta}$ because this observable is computed by an absolute value.

The effective mass $m_{\beta\beta}$ applied to the cases B$_3$ and B$_4$
has the property that, if we do not care that $\theta_{23}$ and $\cos\delta$
are actually determined by equations~(\ref{t23}) and~(\ref{zeta}) and simply
plug in $\theta_{23} = 45^\circ$
and $\cos\delta = 0$, we obtain the equality $m_{\beta\beta} = m_3$. 
Since for a quasi-degenerate neutrino mass spectrum $m_1 \simeq m_3$ holds,
this demonstrates that we should expect
\be
m_{\beta\beta} \simeq m_1
\ee
in the limit of quasi-degeneracy. Numerically it turns out that 
the deviation of $m_{\beta\beta}/m_1$ from one is very small---even at 
$m_1 = 0.05$\,eV, for the inverted spectrum, the ratio $m_{\beta\beta}/m_1$
deviates from one by only $-3.2$\%, at $m_1 = 0.1$\,eV the deviation is 
$-1.5$
per mill. For the normal spectrum this ratio is even closer to one.
This renders a plot $m_{\beta\beta}$ vs.\ $m_1$ superfluous. 
The smallness of $m_{\beta\beta}/m_1 - 1$ is partially explained by the
smallness of $s_{13}^2$ which brings the phases of both $\mu_1$ and $\mu_2$
close to $\pi$~\cite{xing1}. One can check that choosing a large (and thus
unphysical) $s_{13}^2$ there is indeed a substantial deviation of 
$m_{\beta\beta}/m_1$ from one at the lower end of our range of $m_1$.

\section{The remaining cases}
\label{remaining}

Here we will show that the remaining five cases of two texture zeros in $\mnu$
are either such that the assumption of a quasi-degenerate spectrum is
incompatible with the data or that they do not conform to the philosophy put
forward in this paper. 

Cases A$_1$ and A$_2$ are incompatible with quasi-degenerate neutrino
masses, as was noticed in~\cite{FGM}. This can be seen in the
following way. From $\left( \mnu \right)_{ee} = 0$, 
assuming a quasi-degenerate spectrum and using equation~(\ref{V}) we
readily find 
\be
s_{13}^2 \gtrsim c_{13}^2 \left( c_{12}^2 - s_{12}^2 \right) = 
c_{13}^2 \cos(2\theta_{12}),
\ee
in contradiction to our experimental knowledge on $\theta_{13}$ and 
$\theta_{12}$.

Next we consider cases B$_1$ and B$_2$. Taking into account that one knows from
experiment that $s_{13}^2$ is small, in first order in $s_{13}$ for~B$_1$
one obtains~\cite{FGM}
\be
\frac{\mu_1}{m_3} \simeq -\left[ t_{23}^2 + s_{13} 
\left( e^{-i\delta} t_{23} + e^{i\delta}/ t_{23}\right)/t_{12} \right],
\quad
\frac{\mu_2}{m_3} \simeq -\left[ t_{23}^2 - s_{13} 
\left( e^{-i\delta} t_{23} + e^{i\delta}/ t_{23} \right)/t_{12} \right].
\ee
Now we ask the question if the assumption of a quasi-degenerate mass
spectrum compellingly leads to $t_{23} \simeq 1$. The answer is ``no''
because we could choose $s_{13}/(t_{23}t_{12}) \simeq 1$ in order to
achieve quasi-degeneracy; even with the experimentally allowed values for
$s_{13}$ and $t_{12}$ we would obtain a rather small 
$t_{23} \simeq s_{13}/t_{12}$, far from maximal atmospheric neutrino
mixing. Case B$_2$ can be discussed analogously.

Case~C is a bit more involved---for details see~\cite{GL05}. 
In the case of the inverted spectrum, maximal atmospheric neutrino
mixing is not compelling.
For the normal ordering of the spectrum, using the experimental knowledge on
the mass-squared differences and the mixing angles $\theta_{12}$ and
$\theta_{13}$ it follows that $t_{23}$ is extremely close to one and
that the spectrum is quasi-degenerate. However, if we do not use the 
experimental information on $s_{13}^2$, we could assume $c_{13}^2$ being
small instead which would then admit $t_{23}$ being smaller than
one. This is in contrast to cases B$_3$ and B$_4$ where, for
quasi-degeneracy, $t_{23}$ is always close to one, independently of the 
values $s_{12}^2$ and $s_{13}^2$ assume.

\section{Conclusions}
\label{concl}

In this paper we have considered the possibility that neutrinos differ from
charged fermions not only in their Majorana nature but also in a
quasi-degenerate mass spectrum, in stark contrast to the hierarchical mass
spectra of the charged fermions. The appealing aspect of this assumption is
that it is already under scrutiny by present experiments, and more
experiments will join in the near future~\cite{rpp}. Such experiments search for
neutrinoless double-beta decay, whose decay amplitude is proportional to the
effective mass $m_{\beta\beta}$, and for a deviation in the shape of the
endpoint spectrum of the $\beta$-decay of $^3$H 
which is, in essence, sensitive to the average of the 
squares of the neutrino masses if the spectrum is quasi-degenerate.
Moreover, that neutrino mass effects in cosmology have not yet been observed
puts already a stringent although model-dependent bound on the sum of the
masses.

However, the aspect on which we elaborated in this paper
was the possibility to obtain near maximal atmospheric neutrino mixing
from a quasi-degenerate neutrino mass spectrum. The idea is quite simple: if
we have a model with symmetries enforcing a diagonal charged-lepton mass
matrix and the atmospheric neutrino mixing angle being a function of the
neutrino mass ratios, then in the limit of quasi-degeneracy this mixing angle
will become independent of the masses. We have found two instances in the
framework of two texture zeros in the Majorana neutrino mass matrix where 
in this limit atmospheric neutrino mixing becomes maximal, namely the cases
B$_3$ and B$_4$ of~\cite{FGM}. We have shown that these two textures have the
following properties if the neutrino mass spectrum is quasi-degenerate:
\begin{enumerate}
\item
Using the mass-squared differences as input, 
the value of $s_{23}^2$ tends to 1/2 irrespective of the values of
$s_{12}^2$ and $s_{13}^2$; therefore, maximal atmospheric neutrino mixing
has to be considered a true prediction of the textures B$_3$ and B$_4$
in conjunction with quasi-degeneracy.
\item
If $s_{13}^2$ is not exceedingly small, then CP violation in lepton
mixing becomes maximal too, \textit{i.e.}, $\cos\delta$ tends to zero.
\item
Exact vanishing of $s_{13}^2$ is forbidden because this would entail
$\Delta m^2_{21} = 0$, however, values as small as $s_{13}^2 = 0.0001$
are nevertheless possible.
\end{enumerate}

The results for the cases B$_3$ and B$_4$ can be understood in the
following way. With the usual phase convention~(\ref{V}) for the mixing
matrix, in equation~(\ref{gamma}) we have $e^{i\phi}=1$ and, therefore, 
$V_{\mu j} \simeq -V_{\tau j}^*$ for $j=1,2$ and 
$V_{\mu 3} \simeq V_{\tau 3}^*$ for a quasi-degenerate spectrum.
The signs we obtained here are convention-dependent 
and have no physical significance.
That the exact relation $V_{\mu j} = V_{\tau j}^*$ for $j=1,2,3$ 
is a viable and predictive restriction of $V$ 
was already pointed out in~\cite{HS}, later on 
in~\cite{cp} a model was constructed where this relation is enforced by a
generalized CP transformation and softly broken lepton numbers, and  
it was also shown that such a mixing matrix leads to 
$\theta_{23} = 45^\circ$ and $s_{13} \cos\delta = 0$ at the tree level.  
While in~\cite{cp} the symmetry structure is non-abelian and a type 
of $\mu$--$\tau$ interchange symmetry (see~\cite{mu-tau} for some
early references, and~\cite{he} for a recent paper and references
therein), the textures B$_3$ and B$_4$ can be 
enforced by abelian symmetries and, provided the neutrino
mass spectrum is quasi-degenerate, we have the \emph{approximate} relations 
$\theta_{23} \simeq 45^\circ$ and $\cos\delta \simeq 0$.
Therefore, we have shown that maximal atmospheric neutrino mixing
could have an origin completely different from $\mu$--$\tau$
interchange symmetry, it could simply be a consequence of texture
zeros and quasi-degeneracy of the neutrino mass spectrum.

\end{document}